\documentclass{aa}

\usepackage{url}
\usepackage[breaklinks=true]{hyperref}
\usepackage{twoopt}
\usepackage[english]{babel}          
\usepackage{pdflscape}

\usepackage{natbib}
\bibpunct{(}{)}{;}{a}{}{,} 

\usepackage[utf8]{inputenc}
\usepackage[normalem]{ulem}
\usepackage{siunitx}
\usepackage{amsmath, amssymb}
\usepackage[farskip=0pt]{subfig}

\usepackage{multirow,bigstrut,ctable}
\usepackage{rotating}
\usepackage[varg]{txfonts} 
\usepackage{threeparttable}

\hypersetup{
    colorlinks,
    citecolor={blue!50!black},
}

\newcommand{\asec}{\text{$^{\prime\prime}$}}
\newcommand{\calib}{\text{3C 196}}
\newcommand{\trg}{\text{3C 190}}
\newcommand{\Msun}{M$_{\odot}$}
\newcommand{\bbn}{b_{\mathsf{n}}\beta_{\mathsf{n}}}

\begin{document}

\title{The first detection of radio recombination lines at cosmological distances}
\titlerunning{Radio recombination lines at cosmological distances}

\author{K. L. Emig\inst{1}, P. Salas\inst{1}, F. de Gasperin\inst{1,2}, J. B. R. Oonk\inst{1,3}, M. C. Toribio\inst{1}, H. J. A. R\"ottgering\inst{1}, A. G. G. M. Tielens\inst{1}}

\authorrunning{Emig et al.}

\institute{
Leiden Observatory, Leiden University, P.O.Box 9513, NL-2300 RA, Leiden, The Netherlands, \email{emig@strw.leidenuniv.nl}
\and Hamburger Sternwarte, Universit\"at Hamburg, Gojenbergsweg 112, D-21029, Hamburg, Germany
\and ASTRON - the Netherlands Institute for Radio Astronomy, P.O.Box 2, NL-7990 AA, Dwingeloo, the Netherlands
}

\date{Received ... / Accepted ...}

\abstract
{Recombination lines involving high principal quantum numbers ($\mathsf{n} \sim 50-1000$) populate the radio spectrum in large numbers. Low-frequency ($<$ 1 GHz) observations of radio recombination lines (RRLs) primarily from carbon and hydrogen offer a new, if not unique, way to probe cold, largely atomic gas and warm, ionised gas in other galaxies. Furthermore, RRLs can be used to determine the physical state of the emitting regions, such as temperature and density. These properties make RRLs, potentially, a powerful tool of extragalactic interstellar medium (ISM) physics. At low radio frequencies, it is conceivable to detect RRLs out to cosmological distances when illuminated by a strong radio continuum. However, they are extremely faint ($\tau_{\mathrm{peak}} \sim 10^{-3} - 10^{-4}$) and have so far eluded detection outside of the local universe.}
{With observations of the radio quasar \trg\ ($z=1.1946$), we aim to demonstrate that the ISM can be explored out to great distances through low-frequency RRLs.}
{\trg\ was observed with the LOw Frequency ARray (LOFAR) and processed using newly developed techniques for spectral analysis.}
{We report the detection of RRLs in the frequency range 112 MHz -- 163 MHz in the spectrum of \trg. Stacking 13 $\alpha$-transitions with principal quantum numbers $\mathsf{n} = 266-301$, a peak 6$\sigma$ feature of optical depth $\tau_{\mathrm{peak}} = (1.0 \pm 0.2) \times 10^{-3}$ and $\mathrm{FWHM} = 31.2 \pm 8.3$ km s$^{-1}$ was found at $z = 1.124$. This corresponds to a velocity offset of $-9965$ km s$^{-1}$ with respect to the systemic redshift of \trg.}
{We consider three interpretations of the origin of the RRL emission: an intervening dwarf-like galaxy, an active galactic nucleus (AGN) driven outflow, and the inter-galactic medium. We argue that the recombination lines most likely originate in a dwarf-like galaxy ($M \sim 10^{9}$ \Msun) along the line of sight, although we cannot rule out an AGN-driven outflow. We do find the RRLs to be inconsistent with an inter-galactic medium origin. With this detection, we have opened up a new way to study the physical properties of cool, diffuse gas out to cosmological distances.}

\keywords{galaxies: active --- quasars: general --- galaxies: ISM --- radio lines: galaxies }

\maketitle

\section{Introduction} 
\label{sec:intro}

Under typical conditions found in the interstellar medium (ISM), the recombination of electrons with singly ionised atoms can result in appreciable level populations at high principal quantum numbers (e.g. $\mathsf{n} \sim 300$). For these Rydberg atoms, the $\alpha$-transitions (i.e. $\Delta \mathsf{n} = 1$) are so low in energy that they are observable at radio frequencies.  

Radio recombination lines (RRLs) that have transitions stimulated by a radiation field have been observed at frequencies $\nu \lesssim 10$ GHz. The integrated strength of RRLs measured as a function of quantum number, and likewise frequency, is highly dependent upon the physical conditions of the gas. With almost 800 spectral lines (per each element) between 10 MHz and 10 GHz, this trait makes them powerful tools for understanding the physical properties of the medium, such as temperature, density, thermal pressure, and cloud size.

Within our own Galaxy, stimulated RRLs have been found in regions dominated by either hydrogen or carbon emission. In largely ionised gas with temperatures of $T_e \sim 8000$ K and densities of $n_e\sim1$ cm$^{-3}$ \citep[e.g.][]{Anantharamaiah1985b}, hydrogen RRLs are brightest. This gas phase peaks in intensity at frequencies between 250 MHz and 1000 MHz \citep{Roshi2000, Zhao1996, Shaver1978a, Pedlar1978}. Additionally, cold ($T_e \sim 100$ K) yet diffuse ($n_e \sim 0.05$ cm$^{-3}$) gas can result in carbon being highly stimulated \citep{Shaver1975}. This gas phase occurs in the presence of a radiation field capable of ionising  carbon (ionisation potential of 11.3 eV) yet not hydrogen (13.6 eV). With level populations greatly enhanced by dielectronic capture \citep{Watson1980}, carbon RRLs are the most prominent emitters at $\nu \lesssim 250$ MHz \citep{Konovalenko1980, Payne1989, Oonk2017, Salas2018}.

The detection of low-frequency RRLs is greatly aided towards bright radio sources as the intensity of stimulated transitions is proportional to the strength of the radio continuum, unlike the spontaneous transitions of higher frequency recombination lines associated with HII regions \citep[e.g.][]{Zuckerman1974}. This distinction between stimulated and spontaneous transitions is important because it indicates that low-frequency RRLs can be studied out to cosmological distances with bright radio sources.  The advantages of using stimulated recombination lines to study a variety of regimes was quickly realised by \cite{Shaver1978}, including (i) the study of  ionised gas in normal galaxies; (ii) the study of the physical conditions in the nuclei of galaxies, quasars and absorption-line systems; (iii) the assessment of the importance of free-free absorption in the spectra of extragalactic radio sources; and (iv) redshift determination for unidentified radio sources. 

Out of the 15 extragalactic sources that have been detected via RRLs -- all at $\nu > 1$ GHz and from nearby galaxies \citep[e.g. for review see][]{Gordon2002,Roy2008} -- only the spectrum of M 82 shows clear evidence of stimulation-dominated emission \citep{Shaver1978a}. Non-local sources, such as quasars, were searched at 4.8 GHz, but went undetected \citep{Bell1984}. Stifled by instrument capabilities below $\sim$1 GHz, few searches in extragalactic sources have been performed and even fewer outside of the local universe \citep{Churchwell1979}. Although RRLs can be very useful probes of the ISM, they are challenging to detect observationally. Their major obstacle stems from the very low peak optical depths of the lines \citep[i.e. with peak fractional absorption of $10^{-3} - 10^{-4}$;][]{Gordon2002}. 

An important step in extragalactic exploration at low frequencies came with the discovery of carbon RRLs at 56 MHz in the nearby starburst galaxy, M 82 \citep{Morabito2014} using the LOw Frequency ARray \citep[LOFAR;][]{vanHaarlem2013}. The advent of sensitive low-frequency telescopes has reinvigorated the field, as the large fractional bandwidth allows for the detection of many RRL transitions simultaneously and stacking can increase the signal-to-noise ratio by an order of magnitude \citep[e.g.][]{Oonk2015,Salas2017}.

This paper presents the first result of a survey for RRLs at cosmological distances taking advantage of the capabilities afforded by LOFAR with an in-depth study of the $z = 1.1946$ radio quasar 3C190 (see Section~\ref{sec:target}). This has resulted in the detection of RRLs outside of the local universe for the first time. While our aim was to investigate cold clouds  associated with the HI absorbing gas near to \trg, we uncovered something unexpected: RRLs with a large offset in redshift from the radio source.

\begin{figure}
\includegraphics[width=0.47\textwidth]{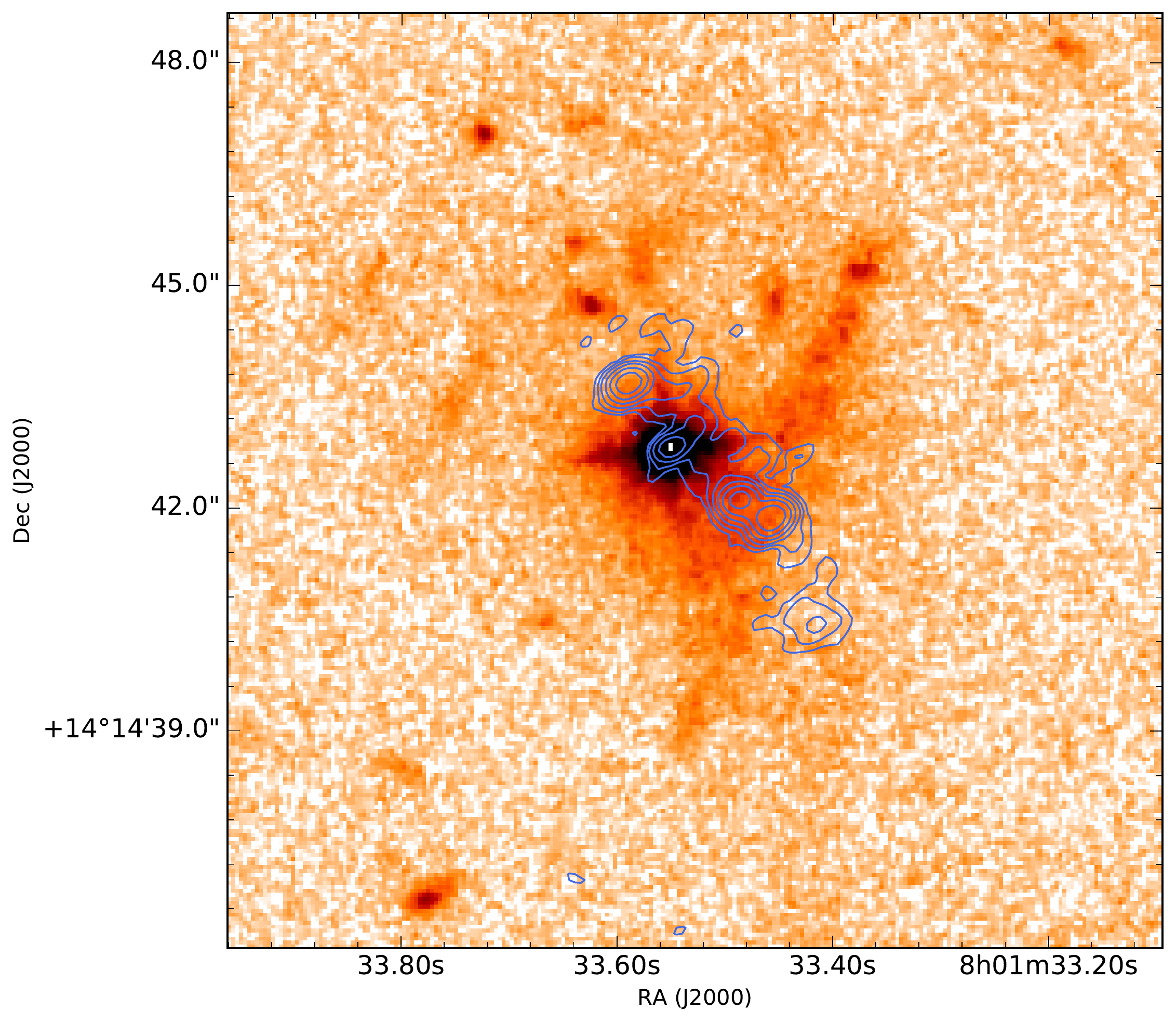}
\caption{Quasar \trg\ (centre) shown in an HST WFPC2 F702W image, where numerous satellite galaxies, a prominent linear feature, and extended diffuse emission can be seen \citep{Stockton2001} associated with the system. Superimposed are MERLIN 1658 MHz radio contours at $\sigma \cdot \log_5([1, 1.5, 2, 2.5, 3, 3.5])$ mJy/beam, showing the 22 kpc extent of the lobe hot spots. While \trg~is a steep spectrum source, it has a flat spectrum core, and thus we expect low-frequency emission to arise in the radio lobes.}
\label{fig:optical}
\end{figure}

\section{Target} 
\label{sec:target}

We identified \trg~as a candidate for RRLs as it is a bright ($\sim 20$ Jy at 140 MHz), steep-spectrum radio galaxy with HI detected in absorption \citep{Ishwara-Chandra2003}. It is classified as a reddened quasar \citep{Smith1980}. Narrow emission lines from [Ne III], [O II], and C III] locate the source at a redshift of $z=1.1946 \pm 0.0005$ \citep{Stockton2001}. As shown in Fig.~\ref{fig:optical}, the host galaxy of \trg\ is the central galaxy of a dense environment that is undergoing several major and minor mergers. Optical spectra reveal the presence of an absorption system at $z=1.19565 \pm 0.00004$ observed via Mg II $\lambda$2798 and Fe II $\lambda \lambda$2343, 2382 and $\lambda \lambda$2586, 2599 \citep{Stockton2001}.

Radio observations at 1662 MHz reveal two hot spots spanning 2.6\asec~\citep{Spencer1991} or 22 kpc in projection. Diffuse emission, stretching to a 4\asec\ extent, indicates the jets may have encountered a dense medium. Using the Giant Metrewave Radio Telescope  (GMRT; resolution of $\sim$5\asec~at 650 MHz), a broad and complex profile of absorbed HI was detected with five Gaussian components spanning almost 600 km s$^{-1}$ \citep{Ishwara-Chandra2003}. Most of these components are blue-shifted with respect to \trg. Of particular interest for carbon RRLs is a deep ($\tau_{\mathrm{peak}} = 0.0100 \pm 0.0003$) and narrow (FWHM $= 66.8 \pm 2.2$ km s$^{-1}$) feature that lies at a velocity offset of $-210.2\pm 1$ km s$^{-1}$. As the region shows evidence of shocked, turbulent gas, the narrow HI likely results from the radio jet interacting with the ambient medium \citep{Ishwara-Chandra2003}.

\begin{figure*}
\centering
\includegraphics[width=0.9\textwidth]{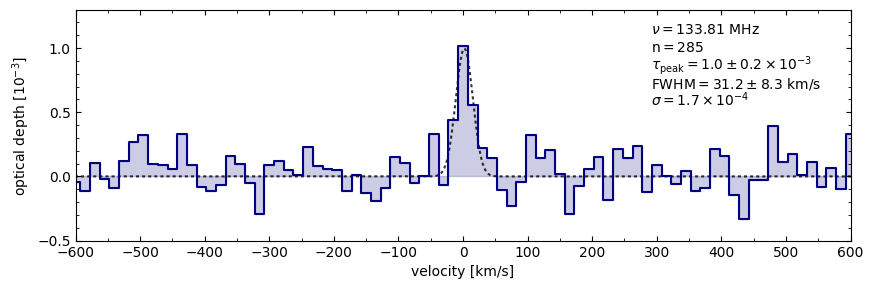}
\caption{Detection of a radio recombination line in emission at quantum level $\mathsf{n}_{\mathrm{eff}} = 285$, with a velocity centred on $z = 1.12405$ ($z = 1.12355$), originating from hydrogen (carbon). This is the average profile of a single line, effectively at 133 MHz, as a result of stacking 13 recombination lines in the spectrum of \trg.}
\label{fig:detection}
\end{figure*}

\section{Observations  and data reduction} 
\label{sec:obs_dr}

\trg\ was observed with the LOFAR High Band Antennas (HBA) on 14 January 2017 (Project ID \texttt{LC7\_027}). Four hours were spent on \trg, with ten minutes on the primary calibrator \calib~before and after. The 34 stations of the Dutch array were used in \texttt{HBA\_DUAL\_INNER} mode. Applying the HBA-low analogue filter, we observed between 109.77 MHz and 189.84 MHz. The observing band is split into sub-bands (SBs) of 195.3125 kHz via a poly-phase filter. After which, each SB is divided into 64 channels and recorded at a frequency resolution of 3.0517 kHz. While data were taken at 1 s time intervals, RFI removal and averaging to 2 s were performed before storing the data.

Processing of the LOFAR data was performed with the SURFSara Grid processing facilities\footnote{https://www.surfsara.nl} \citep[e.g. see][]{Mechev2017, Mechev2018}. While a more detailed description of the data processing can be found in Emig et al. in prep, we summarise the steps below. Starting with the calibrator data, we flagged the first and last four edge channels of each SB, flagged for RFI using \texttt{AOflagger} \citep{Offringa2012}, selected only the core stations (max baseline $\sim$ 4 km), and averaged the data to resolutions of 6 s and 32 channels per SB (or 6.1034 kHz channels). Using \texttt{DPPP} \citep{VanDiepen2018}, we solved for diagonal gain. With \texttt{LoSoTo} \citep{DeGasperin2019} we found the median amplitude solution in time for each channel, creating a per channel bandpass solution. Then, a sixth order polynomial across 10 SBs (2 MHz) was fit to take into account the 1 MHz standing wave, the global slope of the bandpass, and smooth over its scatter (e.g. due to the poly phase filter). These effects are expected to be time independent. 

Next for the target data, we implemented the same flagging steps as for the calibrator and then applied the bandpass solutions. Flagging once more, for each SB we solved for phase only with \texttt{DPPP} on a 6 s time interval, using a LOFAR Global Sky Model \citep{vanHaarlem2013} generated sky model of the field. After averaging the data to a 30 s time resolution, we solved for the amplitude at full-frequency resolution. Importing the solution tables of all SBs into \texttt{LoSoTo}, we flagged for outliers (namely to catch RFI due to digital audio broadcasting with broadband effects) and smoothed the solutions in frequency space, with a running Gaussian of 4 SB FWHM, to ensure that any spectral features are not calibrated out. Once we applied the smoothed amplitude solutions, we imaged each channel with \texttt{WSCLEAN} \citep{Offringa2014}, in which multi-frequency synthesis was used per SB to make a continuum image and extract the clean components for the channel images \citep{Offringa2017a}.

\begin{figure*}
\includegraphics[width=0.52\textwidth]{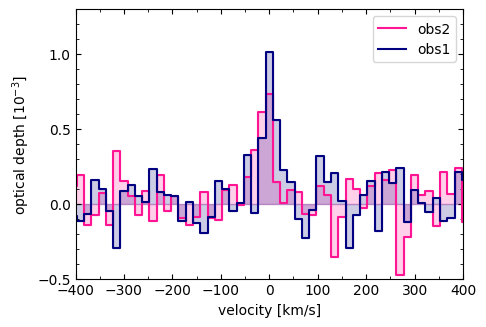}
\includegraphics[width=0.47\textwidth]{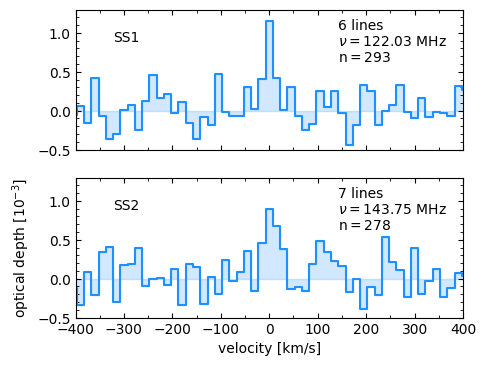}
\caption{\textit{Left:} Observation 1 (blue) as in Fig.~\ref{fig:detection} and the follow-up observation (pink) confirming the feature. These spectra have been Doppler corrected to the LSR. Observation 2, with a factor of 1.6 less time spent on source, is the stack average of 10 lines at quantum number $\mathsf{n}_{\mathrm{eff}} = 287$ and $\nu_{\mathrm{eff}} = 130.9$ MHz. \textit{Right:} We split the spectral lines from Observation 1 into 2 groups, sub-stack 1 (SS1) consisting of the 6 spectral lines with $\mathsf{n} > 285$ and sub-stack 2 (SS2) containing the 7 lines below.}
\label{fig:tests}
\end{figure*}

\section{Spectral stacking}
\label{sec:spec_stack}

\subsection{Spectral processing} 
\label{sec:spec_process}

In this section we describe the processing of the data post calibration and imaging. In summary, once the channel images were created, we convolved them to the same resolution, extracted flux from a fixed aperture, removed the continuum flux, flagged SB spectra, and stacked the spectra in velocity space. 

We first convolved every channel image to the same resolution of 236\asec, a few percent larger than the lowest resolution image, using \texttt{CASA} \citep{McMullin2007}.  The flux density was then extracted from a fixed circular aperture of diameter 236\asec. A spectrum was created for each SB. A fit to the continuum was made to each SB spectrum with a low (1st or 2nd) order polynomial chosen to minimise the chi-square of the fit. For a given redshift, we blanked the channel of the expected frequency of the line when fitting for the continuum. At low frequencies, if stimulated transitions dominate, we have that $I_{\mathrm{line}} \approx I_{\mathrm{cont}} \tau_{\mathrm{line}}$, where the intensity we extract from the observations is $\approx I_{\mathrm{line}} + I_{\mathrm{cont}}$. Therefore, we subtracted the continuum fit and divided by it, which resulted in a measure of the optical depth.  

We inspected spectra on a per SB channel basis. In each spectrum we interpolated over channels which had amplitudes higher than five times the spectrum rms. We interpolated
over channels for which $>$60 \% of the visibility data were flagged as well. 

Examination of spectra on a per-SB basis was done first by eye to catch clear bandpass-related outliers, for example owing to broadband RFI. We also discarded SBs for which their rms or chi-squared of the continuum fit was a 5$\sigma$ outlier (considering a rolling window of 20 SBs centred on the SB in question). Lastly, if $>$40 \% (11) of the channels had been clipped or flagged, we did not include the SB in the stack.

In terms of the lines going into the stack, if one of the clipped channels fell within the blanking region of the expected spectral line, the line was not included in the final stack. Furthermore, we required at least three channels on either side of the blanked region to estimate the continuum, otherwise the line was discarded.

Taking the central frequency of each SB, we determined the spectral line closest in frequency and used it to convert the channels to velocity units using the radio definition of velocity, $ v^{radio} = \frac{\nu_0-\nu}{\nu_0}\,\,c$, where $\nu_0$ is rest frequency.
At this point, we corrected for the velocity offset necessary for Doppler tracking relative to the local standard of rest (LSR) frame. We then interpolated the velocities to a fixed grid with a channel spacing of 15 km s$^{-1}$.

With the spectra aligned in velocity space, the weighted mean optical depth in each channel was found via
\begin{equation}
<\tau_{\mathrm{chan}}> = \frac{ \Sigma_{i=0}^N (w_i \tau_i) }{ \Sigma_{i=0}^N w_i  }
\label{eq:weighted_tau}
,\end{equation}
where $i$ represents each SB going into the stack, and the weighting factor was determined by one over the noise variance of each SB spectrum, $w_i = \sigma_i^{-2}$. The effective frequency and quantum number of the stacked spectrum was determined by the weighted mean values of each SB containing a spectral line.

\subsection{Statistical identification} 
\label{sec:stats_method}

We stacked RRLs across redshift space to search for features within the source and along the line of sight. This was sampled at an interval equivalent to the channel width $v = 15$ km s$^{-1}$ or $\Delta z = 10^{-5}$ and ranged over $z = [0,1.31]$. We note that with the flagging procedure above, the number of spectral lines in each stack at a given redshift did not remain constant. While the following methods are described extensively in Emig et al. in prep., we briefly summarise them here. 

For each redshift tested, we performed a cross-correlation between a template spectrum and the pre-stacked spectra, both in units of optical depth. The template spectrum was populated with Gaussian line profiles at the location of the spectral lines that contributed to the final stack. The line profiles had a peak of one and their full width at half maximum (FWHM) was set by an assumed blanking region. We then took the cross-correlation and normalised it proportionally with the number of lines that went into the stack, i.e. the total area under the template spectrum. This was the same procedure implemented in \cite{Morabito2014}, except we included a normalisation since the number of lines included at each redshift did not remain the same. 

As a second test, we took a template spectrum, stacked the spectral lines at an assumed redshift, and integrated the signal within an assumed FWHM. Furthermore, we stacked and integrated the template spectrum at a range of redshifts, from $z - 0.01$ to $z + 0.01$, at redshift intervals of $10^{-5}$. We then cross-correlated ({\it a}) with ({\it b}): ({\it a}) the integrated optical depth of the template stack as a function of redshift,  and ({\it b}) the observed integrated optical depth of the stacks at each redshift. With this cross-correlation, we corroborate ``mirrors'' of the signal that can be found at a $\Delta z =  \overline{ \Delta \nu_{\mathsf{n},\mathrm{eff}} } / \nu_{\mathsf{n},\mathrm{eff}}$, or multiples thereof, where $\nu_{\mathsf{n},\mathrm{eff}}$ is the frequency of the effective $\mathsf{n}$-level of the stack, and $\overline{ \Delta \nu_{\mathsf{n},\mathrm{eff}}}$ is the average of the change in frequency between $\mathsf{n}_{\mathrm{eff}}$ and all other $\mathsf{n}$ levels included. We digress to explain the aforementioned mirrors. The distance between each recombination line in frequency space is unique, thus allowing us to accurately determine redshift. However, the difference in spacing between $\alpha$-transitions $\mathsf{n}$ and $\mathsf{n} + 1 $ is small ($\sim$1 \%); in other words, the frequencies at which recombination lines fall are close to being, but not quite, periodic. Therefore mirrors of the feature, which are broadened and reduced in peak intensity compared to the original, occur at offsets in redshift that match the frequency spacing between adjacent lines, or more precisely, $\Delta z =  \overline{ \Delta \nu_{\mathsf{n},\mathrm{eff}} } / \nu_{\mathsf{n},\mathrm{eff}}$. It was necessary to include the second  cross-correlation on account of the low signal-to-noise regime of the lines coupled with poor estimation of the continuum over narrow SBs. 

We required that both cross-correlation methods result in a $> 5 \sigma$ value at a redshift to report a detection.

\section{Results}
\label{sec:result}

A significant feature ($> 5\sigma$) was found in the spectrum of \trg\ when considering a line blanking of 15 km s$^{-1}$, arising from the $\alpha$-transitions of hydrogen (carbon) at a redshift of $z=1.12405$ (1.12355) $\pm\ 0.00005$. A redundancy exists between carbon and hydrogen as their $\alpha$-transitions are regularly offset by 149.4 km s$^{-1}$. The stacked feature includes 13 recombination lines of principal quantum numbers $\mathsf{n} = 266-301$, and it has an effective frequency of $\nu_{\mathrm{eff}} = 133.81$ MHz and quantum level of $\mathsf{n}_{\mathrm{eff}} = 285 $. Increasing the blanking region further to 50 km s$^{-1}$, we find the averaged spectral feature (Fig.~\ref{fig:detection}) to have a $31.2 \pm 8.3$ km s$^{-1}$ width, thus an under-sampled Gaussian with our velocity resolution, and an average integrated strength per line of $\int \tau \mathrm{d}\nu = -14.8 \pm$ 7.4 Hz at 6.3$\sigma$. Properties of the line and spectrum are listed in Tab. \ref{tab:detect_params}. 

The per-channel coverage is about two times lower in the blanked region versus the non-blanked. With the weighting described by Eq. \ref{eq:weighted_tau}, the noise outside of the blanked region is effectively $\sqrt{2}$ times lower than the rms reported in Tab. \ref{tab:detect_params}. At $z = 1.12405$, there are a total of 36 RRLs between 112 MHz -- 165 MHz. However, 13 were included in the final stack. There were 22 lines that fell within three channels of the SB edge. An additional spectral line was discarded owing to a poor SB bandpass, identified as an outlier in both its rms and continuum fit.

\begin{table}
\centering
\begin{tabular}{lll}
\hline
$\mathsf{n}_{\mathrm{eff}}$ & 285                   & \\
$\nu_{\mathrm{eff}}$        & 133.81 MHz            & \\
$\tau_{\mathrm{peak}}$      & $(1.0 \pm 0.2) \times 10^{-3}$    & $19 \pm 4$ mJy \\
FWHM                        & $13.9 \pm 3.7$ kHz    & $31.2 \pm 8.3$ km s$^{-1}$ \\
rms                         & $1.7 \times 10^{-4}$  & 3.3 mJy \\
$\int \tau\,\mathrm{d}\nu$  & $-14.8 \pm 7.4$ Hz    & $640 \pm 320$ mJy km s$^{-1}$ \\
\hline
\end{tabular}
\caption{Spectral properties of the stacked RRLs. The values in the right column were determined by considering a measured flux density of $19.3 \pm 3.9$ Jy at 133.81 MHz.}
\label{tab:detect_params}
\end{table}

As further validation, we confirm the RRL detection (1) with a second observation, (2) by performing jack-knife tests \citep{Miller1974}, (3) stacking in two independent sub-groups, and (4) stacking other sources in the field at the same redshift. These tests all give further confidence to the detection and are described below. Additionally, stacking 3C 190 at other redshifts is consistent with noise.

(1) A second observation of \trg~was obtained on 03 May  2017 with the same observational set-up, but with poor ionospheric conditions and thus only 2.5 hours of usable data. Having been taken several months apart, the Doppler correction differs by 32 km s$^{-1}$. As shown on the left side of Fig.~\ref{fig:tests} in pink, a 4.6$\sigma$ feature, including ten spectral lines at $\mathsf{n}_{\mathrm{eff}} = 287$ and $\nu_{\mathrm{eff}} = 130.93$ MHz, is again seen at $z=1.12405$ ($z=1.12355$).

(2) Furthermore, jack-knife tests, as in \cite{Oonk2014}, were performed on the stacked spectrum, in which we iteratively stack the spectrum, each time discarding one line from the stack. The line properties of the stacks do not differ substantially, indicating that a single SB is not responsible for the signal.

(3) We also split the lines into two groups \citep[see also][]{Oonk2014} resulting in two independent stacks, shown in the right-hand side of Fig.~\ref{fig:tests}. Sub-stack 1 (SS1) consists of six spectral lines with $\nu_{\mathrm{eff}} = 122.03$ MHz and $\mathsf{n}_{\mathrm{eff}} = 293$; Sub-stack 2 (SS2) consists of seven lines with $\nu_{\mathrm{eff}} = 143.75$ MHz and $\mathsf{n}_{\mathrm{eff}} = 278$. The line profiles are consistent among the two stacks. The line is more narrow but higher in its peak optical depth in SS1 as compared with SS2, which is consistent with Doppler broadening effects.

(4) Lastly, we stack the next two brightest sources in the field, 3C 191 and 4C +15.22, at $z=1.12405$ as well. While their noise is significantly higher and thus so are their limits on the optical depth, we confirm no prominent feature is seen in their spectra. This gives further confidence that the detection is not an instrumental systematic. 

\begin{figure}
\includegraphics[width=0.48\textwidth]{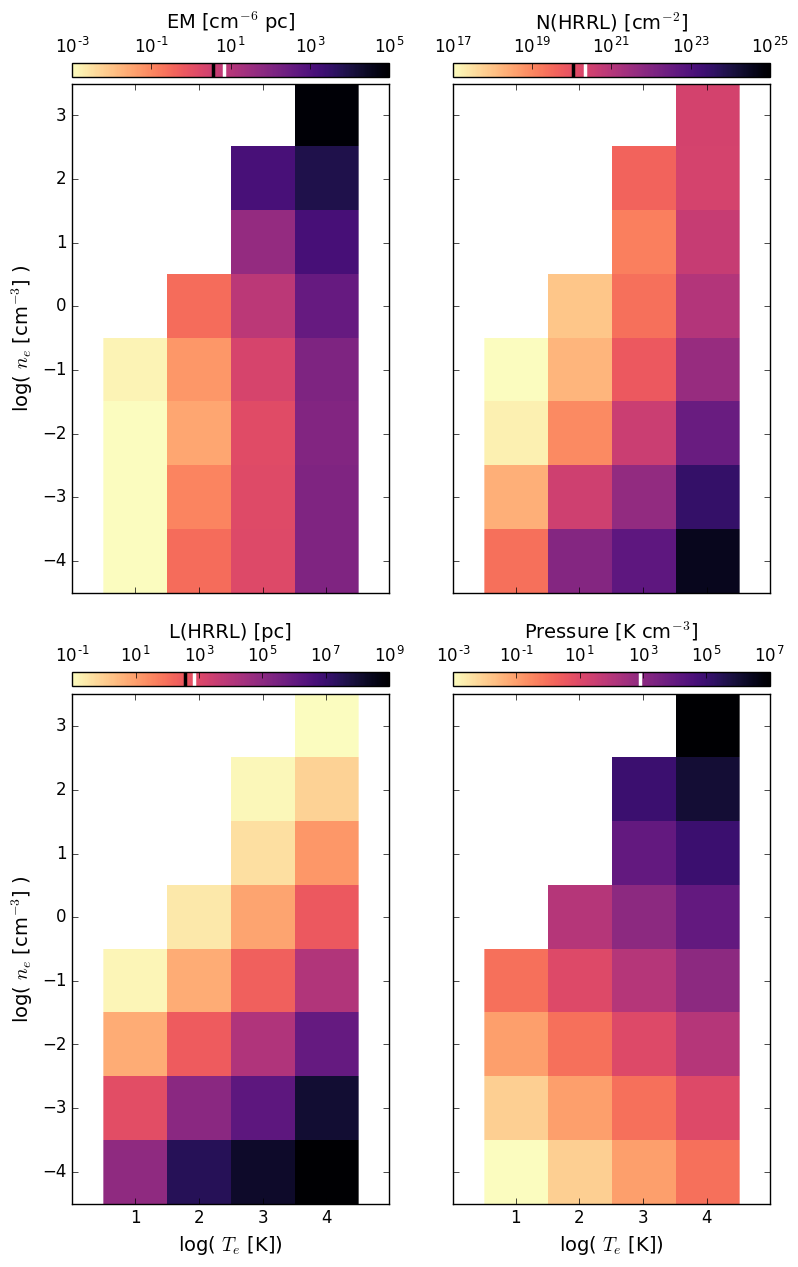}
\caption{Derived physical properties from modelling of \textit{hydrogen} RRL departure coefficients. For input physical conditions, electron density, and temperature, shown on the axes, we computed a departure coefficient ($\bbn$) via RRL modelling. Plugging in the $\bbn$, the integrated optical depth, quantum number, $n_e$ and $T_e$ into Eq.~\ref{eq:inttau}, we derive the emission measure (top left) and corresponding column density (top right) and path length (bottom left) of the RRL emission. Additionally we show the electron pressure (bottom right). Values representing the warm, ionised phase of a face-on (black) and an edge-on (white) Milky Way are shown with dashes in the colour bars. Hydrogen RRLs are expected to be most prominent from warm, $T \sim 1000 - 10\,000$ K, gas.}
\label{fig:H_model}
\end{figure}

\section{RRL Modelling and Interpretation}
\label{sec:model}

RRLs allow us to make constraints on the emitting gas properties arising from the observables of central velocity, line width, and line integrated strength. We discuss the constraints they place on the origin of the detected line. The difference in redshift between \trg\ and the RRL-emitter implies a velocity difference of $-9965$ km s$^{-1}$, corresponding to a (luminosity) distance from \trg\ of 81 Mpc.

To characterise the emitting gas, we use models of radio recombination line emission. Ideally, with multiple RRL detections distributed in frequency, models can be fit to derive gas properties. However, with only one data point, we instead explore a range of input physical conditions (density and temperature) and obtain a range in physical properties (emission measure, column density, path length, and pressure). We use these physical properties to constrain different, possible interpretations on the origin of the emitting gas. 

The integrated optical depth of stimulated recombination lines at low frequencies is described by
\begin{equation}
\int \tau \,\mathrm{d}\nu = 2.046 \times 10^6 \,\mathrm{Hz} \,
\cdot \exp\left(\chi_{\mathsf{n}} \right)
\left( \frac{ T_{\mathrm{e}}}{\mathrm{K}} \right)^{-5/2} 
\frac{EM}{ \mathrm{cm^{-6} \,pc} }
b_{\mathsf{n}}\beta_{\mathsf{n}}
\label{eq:inttau}
\end{equation}
for $\alpha$ transitions \citep[e.g.][]{Shaver1975,Salgado2017a}. Here $\chi_{\mathsf{n}} = 1.58\times 10^5 \,\mathrm{K} / ( \mathsf{n}^2 T_{\mathrm{e}})$, $\mathsf{n}$ is the quantum level, and $EM$ is the emission measure expressed as $EM = n_e n_{\mathrm{ion}} L_{\mathrm{ion}}$ for electron density $n_e$, ion density $n_{\mathrm{ion}}$, and path length $L_{\mathrm{ion}}$. The coefficients that describe the gas departure from local thermodynamic equilibrium, $b_{\mathsf{n}}\beta_{\mathsf{n}}$, require detailed modelling of the atomic physics, for which we refer to the comprehensive low-frequency models of \cite{Salgado2017a, Salgado2017b}.

We explored models with electron densities ranging from $10^{-4}$ cm$^{-3}$ -- $10^{3}$ cm$^{-3}$, electron temperatures of $10$ K -- $10^{4}$ K, and with four different radiation fields. We wanted to cover a wide range of parameters that include the typical temperatures of carbon (10 K -- 100 K) \citep{Oonk2017, Salas2017} and of hydrogen ($10^3$ K -- $10^4$ K) \citep{Anantharamaiah1985b,  Heiles1996} RRL emitting gas. We note, the assumptions made for dielectronic capture in \cite{Salgado2017a} is relevant only for $T < 15\,000$ K \citep{Watson1980}. The densities were chosen to cover typical parameters found within our Galaxy in diffuse clouds, HII region outskirts, and extended low-density phases of the ISM \citep[e.g.][]{Tielens2005, Ferriere2001}. Moreover, we incorporated more extreme conditions which resemble those of the (less dense) inter-galactic medium (IGM) and (possibly, more dense) AGN environments.  The radiation fields we consider have (1) only cosmic microwave background (CMB) radiation (for $z = 0$), (2) CMB radiation and a synchrotron field with $\beta = -2.6$ \citep[e.g.][]{deOliveira-Costa2008, Klein2018} and scaled to $T_r = 800$ K at 100 MHz, (3) to $T_r = 2000$ K, and (4) to $T_r = 10^5$ K. For the 19.3 Jy flux density of \trg\ at 133.81 MHz, the synchrotron brightness temperature is given by $T_b = (1+z) \, 4.7 \times 10^{4}$ K sr$^{-1}$. For an absorber 81 Mpc away with an estimated $\Omega \sim 2 \times 10^{-7}$ sr, the effective radiation temperature from \trg~is negligible ($T_r = 0.02$ K). This implies that if the RRLs originate in an intervening absorber, the continuum from \trg\ does not contribute significantly to $T_{r}$. On the other hand, material that is $\sim$15 kpc away (or at least within a Mpc) would see a significantly higher radiation temperature, $T_r \sim 10^5$ K. 

\begin{figure}
\includegraphics[width=0.48\textwidth]{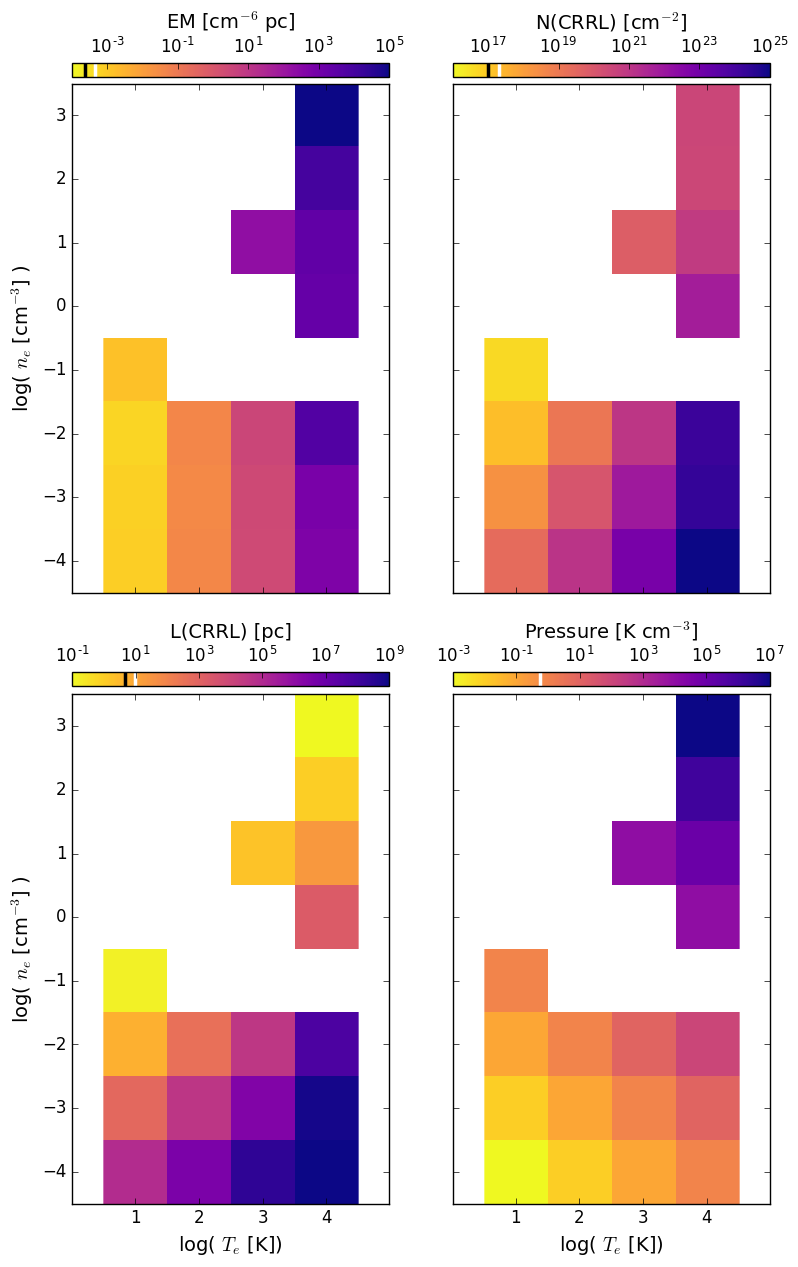}
\caption{Same physical properties as in Fig.~\ref{fig:H_model}, except for \textit{carbon} RRL modelling of the departure coefficients and the Milky Way values are representative of its cold, neutral medium. Carbon RRLs are expected to be most prominent from cold, $T \sim 10 - 100$ K, gas.}
\label{fig:C_model}
\end{figure}

Starting from Eq.~\ref{eq:inttau}, we can solve for $EM$ by inputting observed quantities (i.e. $\int \tau \,\mathrm{d}\nu = -14.8 \pm $ 7.4 Hz at $\mathsf{n}_{\mathrm{eff}} = 285$) and the computed departure coefficient and assumed temperature. The physical properties obtained are shown in Fig.~\ref{fig:H_model} and Fig.~\ref{fig:C_model} for hydrogen and carbon, respectively. We assume all ionised electrons originate from their RRL element, and ionised fractions of $X_e = 1$ and  $X_e = A_{\mathrm{C}} = 1.4 \times 10^{-4}$ \citep{Sofia2004} in the respective cases. 
The figures show values obtained for $T_r = 2000$ K; these are representative of the radiation fields considered. The properties derived for the three coolest radiation fields do not differ by more than $\sim$20\%. For the radiation field of $T_r = 10^5$ K, the results are qualitatively similar; they can be found in the Appendix, Fig.~\ref{fig:Tr_1d5}.
We indicate values that the Milky Way would be observed to have if placed at $z=1.124$; we assume cold gas extends to $R = 25$ kpc \citep{Dickey2009} with parameters of the warm, ionised medium, and cold, neutral medium found in \cite{Tielens2005}. We also assume that the background continuum is emitted from a region 4\arcsec\ in extent, equivalent to 33.3 kpc at the redshift of \trg. 

We consider five scenarios to explain the RRLs: (i) hydrogen or (ii) carbon RRLs from an intervening galaxy, (iii) hydrogen or (iv) carbon RRLs from the AGN outflow, and (v) hydrogen RRLs from the IGM. It is not immediately clear whether the RRL originates from hydrogen or carbon because of two main issues. Firstly, a regular separation of 149.4 km s$^{-1}$ exists for the Bohr-like $\alpha$-transitions between the two elements. Secondly, at the frequency at which we detected the feature, rest-frame 284 MHz, cold (carbon origin) and warm (hydrogen origin) gas phases have both been observed in RRLs within our own Galaxy \citep[e.g.][]{Anantharamaiah1985b}. In the following sections, we give a brief description of each scenario and its RRL modelling constraints.

\begin{table*}
\centering
\begin{tabular}{lccc | cccc | cccc}
\hline
Scenario	&$n_e$	&$T_e$	&$P / k$	&$\bbn$	&$EM$			&$N(RRL)$		&$L$		&$\phi$	&$M$        &$Q$        &SFR\\
	&[cm$^{-3}$] 	&[K]	&[K cm$^{-3}$] &	&[cm$^{-6}$ pc] &[cm$^{-2}$] 	&[pc]		&       &[\Msun]    &[phot s$^{-1}$] &[\Msun/yr] \\
\hline
i) galaxy, hydrogen \\
	&1 			&$10^3$ 	&$10^3$ 	&-28.9 	&7.8 		&$2\times10^{19}$ 	&7.8 		&0.10	&$2\times10^{8}$ &$3.6\times 10^{53}$ &3.9 \\
	&1 			&$10^4$ 	&$10^4$		&-205	&350		&$1\times10^{21}$ 	&350 	    &0.10	&$8\times10^{9}$ &$2.5\times 10^{54}$ &27 \\
	&0.1 		&$10^3$ 	&$10^2$		&-103	&2.2		&$7\times10^{19}$ 	&220 	    &0.10	&$5\times10^{8}$ &$1.0\times 10^{53}$ &1.1 \\
	&0.1 		&$10^4$ 	&$10^3$		&-587	&120		&$4\times10^{21}$ 	&$12\,000$  &0.66 &$^*3\times10^{10}$ &$1.9\times 10^{54}$ &21 \\
	&0.01		&$10^3$	    &$10$		&-159	&1.4		&$4\times10^{20}$ 	&$14\,000$  &0.73 &$^*3\times10^{9}$ &$1.4\times 10^{53}$ &1.5 \\
ii) galaxy, carbon \\
	&0.1 		&10 		&$10^{3.9}$	&-1.31 	&0.001 	    &$4\times10^{16}$ 	&0.14 	    &0.10	&$2\times10^{9}$ & & \\ 
	&0.01 		&10 		&$10^{2.9}$	&-3.37 	&0.0006 	&$2\times10^{17}$ 	&5.5 	    &0.10	&$9\times10^{9}$ & & \\ 
iii) outflow, hydrogen \\
	&$10$		&$10^3$	&$10^4$		&-5.23	&44			    &$1\times10^{19}$ 	&0.43 	    &0.0007 &$^*9\times10^{7}$ & & \\
	&$10$		&$10^4$	&$10^5$		&-46.5	&1600		    &$5\times10^{20}$ 	&16 		&0.008	&$^*3\times10^{9}$ & & \\	
iv) outflow, carbon \\
	&0.1		&$10$	&$10^{3.9}$	&-0.76	&0.002 	        &$8\times10^{16}$ 	&0.25 	    &0.0005	&$^*4\times10^9$ & & \\
\hline
\end{tabular}
\caption{Model results for which only \textit{plausible} interpretations are listed. $n_e$, $T_e$: input electron density and electron temperature. For scenario (i) and (ii), a radiation temperature of $T_r = 800$ K was considered, and for scenario (iii) $T_r = 10^5$ K was considered. $P$: the thermal pressure of the input temperature and density, where $P/k \sim (n_e /X_e) T_e$. $\bbn$: the departure coefficient derived from the models. $EM$: the RRL emission measure derived from the model output and Eq.~\ref{eq:inttau}. $N$, $L$: the RRL column density and path length corresponding to the $EM$. $\phi$: the surface filling factor of the object in the beam. $M$: the total mass of a disk-like, face-on galaxy of $R = 5.5$ kpc. $Q$: ionization rate needed to maintain ionized gas of the input $n_e$, $T_e$. SFR: star-formation rate determined from $Q$.\\
$^*$ assume a spherical cloud in calculating the mass. }
\label{tab:model}
\end{table*}

\subsection{Intervening, dwarf-like galaxy}

The RRLs would originate from the disc or in extended material of a dwarf-like galaxy along the line of sight. The galaxy should be small in mass and size such that it is not visible in Hubble Space Telescope (HST) imaging shown in Fig.~\ref{fig:optical} ($M_U > 27.2$). The narrow line-width of the RRL indicates a face-on orientation  consistent with cool phases of the ISM for small galaxies, whether they be from the disc or from a halo component \citep{Tumlinson2013}. In this object, star formation would be possible and thus reasonable for a stimulated radiation field. Since we do not find a counterpart for the RRL in existing spectroscopic observations of the quasar \citep{Stockton2001}, the RRL source should overlap with only the radio emission and not the optical. Furthermore, the high redshift and small mass suggests that the galaxy would be low in metallicity. 

\subsubsection{Hydrogen RRLs in an intervening galaxy}

In scenario (i), the RRL would originate from hydrogen and thus largely ionised gas, either in the disc or in extended material of a dwarf-type galaxy. 

We find sensible results (see Fig.~\ref{fig:H_model}) for models with temperatures $ 10^{2} < T_e / \mathrm{K} < 10^5$ and densities $10^{-2} < n_e /\mathrm{ cm^{-3}} < 10$, in agreement with RRLs observed in the disc of the Milky Way \citep{Anantharamaiah1985b, Heiles1996}. The model results for plausible interpretations of the RRL-emitting gas are listed in Table \ref{tab:model}. 
We list the model input and derived physical properties for models with $T_r = 800$ K. Also listed is an estimate of the mass of warm ionised material; we assume a galaxy of $R = 5.5$ kpc and thus a surface filling factor of $\phi = 0.1$.  We calculate the mass for a face-on disc as $ M = m_H n_H \pi R^2 h$, where $h$ is the scale height of the phase, $h = L / \phi$, and $L$ is the path length derived from the model.

We also make note of warm, ionised gas that could originate in the halo or circum-galactic medium of a small galaxy. The densities would be slightly lower, closer to $n_e  \leq 10^{-2}$ cm$^{-3}$ \citep{Tumlinson2017}, and the path lengths longer, as the gas is distributed outside of the disc. Possibilities for this are also listed in Table~\ref{tab:model}, except we consider this material to be spherically distributed for the mass estimate. For spherical geometry, we let $R^3 = 3/4 L \cdot (16.6 \,\mathrm{kpc})^2$ to calculate the implied true radius of the clump, and extract the beam filing factor. 

It is a relevant exercise to place an upper limit on the temperature of the gas assuming (non-)thermal motions Doppler-broaden the line width.  As defined in \cite{Brocklehurst1972}, a Doppler-broadened line with FWHM ($ \Delta v$) is given by
 $ \Delta v = 30.25 \, \mathrm{km\ s^{-1}} \left( \frac{ m_p }{ m } \frac{ T }{ 2 \times 10^{4} \,\mathrm{K} } \right)^{1/2} $,
where $m_p$ is the proton mass, $m$ is the nuclear mass, and in this work $T$ we take as the electron temperature. We find an upper limit for hydrogen gas of $T_e = 21\,300$ K.

As we find a broad range of physical conditions ($ 10^{2} < T_e / \mathrm{K} < 10^{4.3}$ and $10^{-2} < n_e /\mathrm{ cm^{-3}} < 10$) that could be interpreted as hydrogen RRLs in an intervening, dwarf galaxy, we estimate the number of ionising photons needed to maintain this gas phase.  As described in \cite{Rubin1968}, the ionisation rate, $Q$, needed to maintain the implied physical conditions is given by $Q$ [photons s$^{-1}$]  $= 4.1 \times 10^{-10} n_e n_{\mathrm{ion}} V T_e^{-0.8}$ in a total volume $V$. The computed value of each plausible interpretation is shown in Table \ref{tab:model}. Moreover, we calculate star formation rates (SFRs) from these ionisation rates, via SFR [\Msun\ yr$^{-1}$] $= Q \cdot 1.08 \times 10^{-53}$ \citep{Kennicutt1998a}.

\subsubsection{Carbon RRLs in an intervening galaxy}
In scenario (ii), RRLs from carbon would arise in cold, diffuse gas clouds within the mid-plane of a galaxy. With conservative constraints on thermal pressure, $P < 10^{5}$ K cm$^{-3}$ \citep{Jenkins2001, Herrera-Camus2017}, and on the column density, $N$(CRRL)$<10^{18}$ cm$^{-2}$ such that $N$(HI)$< 10^{22}$ cm$^{-2}$ for a carbon abundance of $A_C = 1.4 \times 10^{-4}$, we rule out many of the possibilities considered in Fig.~\ref{fig:C_model}. However, we find two plausible sets of physical conditions for this scenario, placing limits on electron density of $10^{-3} < n_e / \mathrm{cm}^{-3} < 1 $ and electron temperature of $T_e < 100$ K. We list the results in Table~\ref{tab:model}, along with an estimate of the mass of cold neutral material as described above for the face-on disc, letting $n_H = n_e / A_C$. 

Since this is a viable scenario, we derive integrated SFRs from the mass estimate of the cold, atomic gas. Referencing the relation \cite{Lopez-Sanchez2018} have found for local volume dwarf and spiral galaxies that have gas fractions dominated by HI, we find the estimated masses of $2 \times 10^9$ \Msun\ and $9 \times 10^9$ \Msun\ to have SFRs of $\sim 0.1$ \Msun/yr and $1$ \Msun/yr, respectively. 

\subsection{AGN-driven outflow}
In this scenario, RRLs would originate in gas outflowing as a result of the quasar or of the jet impacting the medium, for instance the optical linear feature apparent in Fig.~\ref{fig:optical}. 

There are notable reasons why this interpretation is less likely. In this system, nothing in the present literature indicates gas is moving at velocities higher than $\sim$600 km s$^{-1}$ \citep{Stockton2001}, one to two orders of magnitude below that of the RRL velocity.  Furthermore, the narrow line width of the RRL is hard to maintain with a very high velocity. 

Despite these indications, we investigated the scenario based on the following reasons. Jet interactions causing cold and ionised gas to outflow up to $\sim$1000 km s$^{-1}$ have been observed in a number of AGN \citep[e.g.][]{Morganti2005}. Via Sloan Digital Sky Survey (SDSS) optical spectra of $\sim$17\,000 quasars, the majority of associated absorbers are found out to 2000 km s$^{-1}$ -- 4000 km s$^{-1}$, with tails to 10\,000 km s$^{-1}$, albeit for warmer gas \citep{Chen2017}. Additionally, warm gas with narrow line widths and velocities of -14\,050 km s$^{-1}$ has been observed in radiation-driven outflows \citep{Hamann2011}, and with velocities up to -3000 km s$^{-1}$ in low-ionisation species, 900 pc away from the quasar \citep{Xu2018}. 

In the case of \trg, its jet speed is estimated to be 0.22c, even out to its current $\sim$15 kpc scale. This is based on orientation \citep{Best1995} and electron ageing \citep{KatzStone1997}. The jet indeed appears to be interacting with the optical linear feature seen through the diffuse radio emission surrounding the hot spots \citep{KatzStone1997} as well as displaced [O II] and a resulting cavity-like feature \citep{Stockton2001}. Again, we point out that the displaced [O II] has a FWHM of 85 km s$^{-1}$ and reaches velocities of -600 km s$^{-1}$ relative to 3C 190.

Two possible outflow scenarios could be (1)  that the RRL originates in ablated [OII] material and is carried to high velocities, approaching that of the jet or wind speeds, and (2)  that the RRL cools out of the shock heated material, and as it approaches $T\sim10^4$ K, it condenses. 

For a 500 pc cloud with a velocity dispersion of 31 km s$^{-1}$, the dissipation time of the cloud is $10^{7}$ yrs. With a bulk motion of 10\,000 km s$^{-1}$, this gas could reach $\sim$100 kpc before dissipating. This implies that observing this type of gas cloud is conceivable.

\subsubsection{Hydrogen RRLs in an AGN-driven outflow}

In this scenario (iii), hydrogen RRLs would originate in (partially) ionised gas. With the higher densities we explore, collisional broadening may cause significant line broadening. We adopt the following expression for the Lorentzian FWHM due to collisions with electrons for $\alpha$-transitions \citep{Salgado2017b},
\begin{equation}
\Delta\nu_{\rm{col}} \approx \frac{n_e}{\rm{cm}^{-3} } \left( \frac{ 10^{a} \mathsf{n}^{\gamma_{\rm{col}}} }{\pi} \right) \rm{Hz},
\end{equation}
where $a$ and $\gamma_{\rm{col}}$ depend on the gas temperature (values for which be found in \cite{Salgado2017b}) and $\sf{n}$ is the quantum level. We use this prescription to place an upper limit on the density of $n_e < 15$ cm$^{-3}$ for our range of temperatures. For a density of $n_e = 10$ cm$^{-3}$ and temperatures of $T_e = 10^3$ K -- $10^4$ K, the thermal pressure would be elevated compared to typical ISM values (see Fig.~\ref{fig:H_model}). High pressures have been found in various gas phases of outflowing material \citep[e.g.][]{Santoro2018,Oosterloo2017,Holt2011}. We list physical conditions that would indeed be possible within this scenario in Table~\ref{tab:model}.

\subsubsection{Carbon RRLs in an AGN-driven outflow}
In scenario (iv), RRLs would originate in an outflow, but from material colder and more dense than the previous scenario.  As we stated in the intervening galaxy example, we expect $N$(CRRL) $ < 10^{18}$ cm$^{-2}$, since the corresponding neutral hydrogen column density limit is $10^{22}$ cm$^{-2}$. We note that for gas with  $T_e < 100$ K and density of $ 0.01 < n_e / \mathrm{cm}^{-3} < 1$, the path length derived from the model is $L = 0.25$ pc. Assuming that the gas is a sphere, this path length would imply a radius of 370 pc and hence a surface filling factor of $5 \times 10^{-4}$. These and other physical properties listed in Table~\ref{tab:model} do not seem unreasonable. We conclude that this scenario could explain the RRLs we observe.

\subsection{Hydrogen RRLs from the intervening IGM} 
We consider gas with typical properties of an isolated IGM cloud at $z\sim1$, with temperature $T_e \sim 10^{4.5}$ K and densities $n_e < 10^{-4}$ cm$^{-3}$ \citep{McQuinn2016}. To test this scenario (v), we extended the RRL models down to densities of $10^{-4}$ cm$^{-3}$, but notice that at low densities and high temperatures, we find unreasonably large path lengths, approaching $10^5$ to $10^7$ kpc, to match the observed feature (see Fig.~\ref{fig:H_model}). Ultimately, we determine it is unfeasible to reproduce the observed integrated optical depth of the RRL with an IGM cloud.

\section{Conclusions}
\label{sec:conclude}

Using LOFAR, we have identified RRLs centred at 133.8 MHz in the spectrum of \trg, as a result of stacking 13 $\alpha$-transitions at $z = 1.124$ (Figure \ref{fig:detection}). This is the first detection of RRLs outside of the local universe.

At low frequencies, recombination lines can occur in diffuse gas from stimulated transitions in hydrogen and carbon. Since their transitions are regularly spaced $\sim$150 km s$^{-1}$ apart, an ambiguity exists in determining the species of origin. However, carbon and hydrogen RRLs originate in distinctly different types of gas. Carbon arises in cold, neutral gas clouds, and hydrogen RRLs arise in warm, largely ionised material. 

In this paper, we demonstrate how RRLs can be used to study the physical properties in these types of gas clouds. We model the non-LTE effects responsible for strong stimulation; inputting physical conditions (temperature and density), we constrain the origin of the gas (Table.~\ref{tab:model}). We find the RRL could be explained by hydrogen ($ 10^{2} < T_e / \mathrm{K} < 10^{4.3}$ and $10^{-2} < n_e /\mathrm{ cm^{-3}} < 10$) or carbon ($T_e < 10^2$ K and $10^{-3} < n_e / \mathrm{cm}^{-3} < 1$) emission in an intervening, dwarf galaxy ($M\sim10^9$ \Msun), roughly 80 Mpc from \trg.  Although we consider it to be less likely, we cannot rule out the possibility that the RRL emitter is outflowing from the radio-loud quasar at $\sim$10\,000 km s$^{-1}$, from either hydrogen or carbon emission. Lastly, we rule out an IGM origin.

Since the RRLs in the spectrum of \trg\ have no counterpart in existing observations, follow-up investigations are crucial. Observations targeting RRLs at a higher and a lower frequency give the best indication of success; they would distinguish between the carbon and hydrogen RRL origin and further constrain the physical conditions of the gas. An intervening galaxy seen only against the radio emission may reveal itself, slightly offset from the quasar, as an absorber in integral-field-spectroscopy observations taken with high-spatial resolution. If the RRL originates from cold, neutral gas, HI 21cm absorption may be found at the redshifted frequency. HI has been searched for over only a small velocity interval centred on the systemic velocity and gas at $\sim$10\,000 km s$^{-1}$ would have been missed. 

The detection of RRLs in the spectrum of \trg\ has opened up a new way to study the physical properties of the ISM out to cosmological distances. The LOFAR Two Metre Sky Survey \citep{Shimwell2017, Shimwell2019} of the Northern Hemisphere is uniquely suited to characterise RRL emitters in a large population of sources. 

\begin{acknowledgements}
The authors would like to thank the referee for constructive and positive feedback. The authors would like to thank Alan Stockton, Anita Richards, and Ralph Spencer for their help in acquiring supplementary data.  We also thank Aayush Saxena, Madusha Gunawardhana, Rogier Windhorst, Turgay Culgar, and Nastasha Wijers for useful discussions. KLE, PS, JBRO, HJAR, and AGGMT acknowledge financial support from the Netherlands Organization for Scientific Research (NWO) through TOP grant 614.001.351. AGGMT acknowledges support through the Spinoza premier of the NWO. MCT acknowledges financial support from the NWO through funding of Allegro. FdG is supported by the VENI research programme with project number 639.041.542, which is financed by the NWO. Part of this work was carried out on the Dutch national e-infrastructure with the support of the SURF Cooperative through grant e-infra 160022 \& 160152. This paper is based (in part) on results obtained with International LOFAR Telescope (ILT) equipment under project code \texttt{LC7\_027}. LOFAR \citep{vanHaarlem2013} is the LOw Frequency ARray designed and constructed by ASTRON. It has observing, data processing, and data storage facilities in several countries, which are owned by various parties (each with their own funding sources) and are collectively operated by the ILT foundation under a joint scientific policy. The ILT resources have benefitted from the following recent major funding sources: CNRS-INSU, Observatoire de Paris and Universite d'Orleans, France; BMBF, MIWF-NRW, MPG, Germany; Science Foundation Ireland (SFI), Department of Business, Enterprise and Innovation (DBEI), Ireland; NWO, The Netherlands; The Science and Technology Facilities Council, UK; and the Ministry of Science and Higher Education, Poland. KLE would like to welcome Zada Gray Emig Tibbits into this world. 

Software:
APLpy \citep{Robitaille2012}, 
astropy \citep{TheAstropyCollaboration2018}, 
CASA \citep{McMullin2007}, 
CRRLpy \citep{Salas2016},
DPPP \citep{VanDiepen2018},
iPython \citep{Perez2007},  
LoSoTo \citep{DeGasperin2019},
matplotlib \citep{Hunter2007}, 
WSCLEAN \citep{Offringa2014}
\end{acknowledgements}

\bibliographystyle{aa}
\bibliography{thesis1}

\newpage
\onecolumn

\begin{appendix}

\section{RRL Modelling for high radiation temperatures}

\begin{figure*}[h!]
\includegraphics[width=0.49\textwidth]{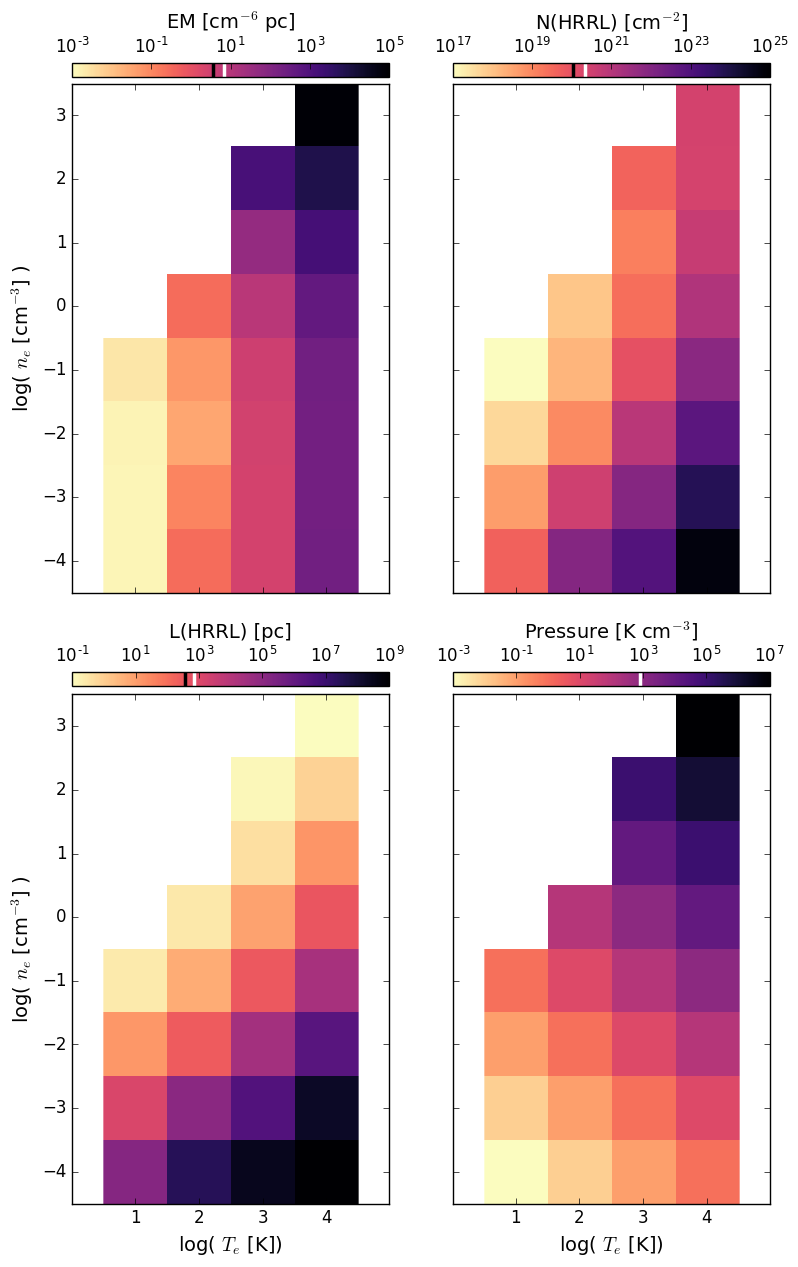}
\includegraphics[width=0.49\textwidth]{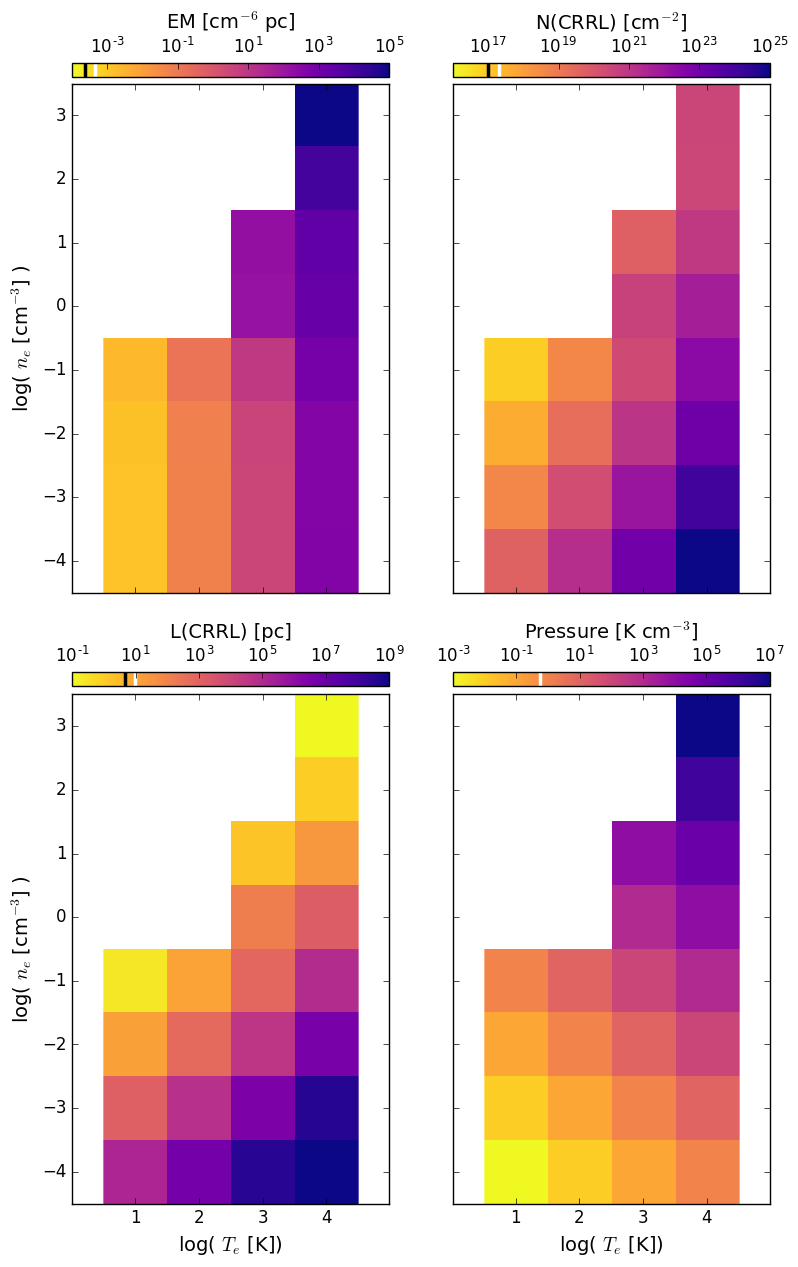}
\caption{Same physical properties as in Fig. \ref{fig:H_model} and \ref{fig:C_model}, except these results for hydrogen (left) and carbon (right) have been derived with $T_r = 10^5$ K.}
\label{fig:Tr_1d5}
\end{figure*}

\end{appendix}

\end{document}